\documentclass[conference]{IEEEtran}
\usepackage[pdftex]{graphicx}
\usepackage{url}
\RequirePackage[colorlinks,citecolor=blue,urlcolor=blue]{hyperref}

\usepackage{amsmath, amsfonts, amssymb, amsthm,color}
\newtheorem{theorem}{Theorem}
\newtheorem{lemma}{Lemma}

\usepackage[font=footnotesize]{subfig}
\usepackage{cite}

\begin{document}
%
\title{High-Dimensional Matched Subspace Detection  When Data are Missing}


\author{Laura Balzano, Benjamin Recht, and Robert Nowak \\
University of Wisconsin-Madison}
\def \E{\mathbb E}
\def \1{\mathbf 1}
\def \P{\mathrm{Pr}}
\def \R{\mathbb{R}}
\def \S{\mathcal{S}}
\def \N{p}

\maketitle

\begin{abstract}
We consider the problem of deciding whether a highly incomplete signal lies within a given subspace.  This problem, Matched Subspace Detection, is a classical, well-studied problem when the signal is completely observed. High-dimensional testing problems in which it may be prohibitive or impossible to obtain a complete observation motivate this work.  The signal is represented as a vector in $\R^n$, but we only observe $m\ll n$ of its elements. We show that reliable detection is possible, under mild incoherence conditions, as long as $m$ is slightly greater than the dimension of the subspace in question.
\end{abstract}


%
\IEEEpeerreviewmaketitle

%
%
\section{Introduction}

Testing whether a signal lies within a subspace is a problem arising in a wide range of applications including medical~\cite{medimag} and hyperspectral~\cite{hyper} imaging, communications~\cite{multiaccess}, radar~\cite{radar}, and anomaly detection~\cite{hyper2}.  The classical formulation of this problem is a binary hypothesis test of the following form.  Let $v \in \R^n$ denote a signal and let $x = v + w$, where $w$ is a noise of known distribution.  We are given a subspace $S \subset \R^n$ and we wish to decide if $v \in S$ or not, based on $x$.  Tests are usually based on some measure of the energy of $x$ in the subspace $S$, and these `matched subspace detectors' enjoy  optimal properties \cite{scharf, scharfpap}.

This paper considers a variation on this classical problem, motivated by high-dimensional applications where it is prohibitive or impossible to measure $v$ completely.  We assume that only a small subset $\Omega \subset \{1,\dots,n\}$ of the elements of $v$ are observed (with or without noise), and based on these observations we want to test whether $v \in S$.  For example, consider monitoring a large networked system such as a portion
of the Internet. Measurement nodes in the network may have software that
collects measurements such as upload and download rate, number of packets,
or type of traffic given by the packet headers. In order to monitor the
network these measurements will be collected in a central place for compilation,
modeling and analysis. The effective dimension of the state of such systems
is often much lower than the extrinsic dimension of the network itself.
Subspace detection, therefore, can be a useful tool for detecting changes or
anomalies. The challenge is that it may be impossible to obtain every
measurement from every point in the network due to resource constraints,
node outages, etc.

The main result of this paper answers the following question.  Given a subspace $S$ of dimension $r\ll n$, how many elements of $v$ must be observed so that we can reliably decide if it belongs to $S$?  The answer is that, under some mild incoherence conditions,
the number is $O(r \log r)$.  This means that reliable matched subspace detectors can be constructed from very few measurements, making them scalable and applicable to large-scale testing problems.

The main focus of this paper is an estimator of the energy of $v$ in $S$ based on only observing the elements $\{v_i\}_{i \in \Omega}$.  Section~\ref{sec:estimator} proposes the estimator.  Section~\ref{sec:main} presents a theorem giving quantitative bounds on the estimator's performance and the proof using three lemmas that are proved in the Appendix. Section~\ref{sec:exp} presents numerical experiments. Section~\ref{sec:detect} applies the
main result to the subspace detection problem, both with and without noise.

\section{Energy Estimation from Incomplete Data}
\label{sec:estimator}

Let $v_\Omega$ be the vector of dimension $|\Omega|\times 1$ comprised of the elements $v_{i}$, $i\in \Omega$, ordered lexigraphically; here $|\Omega|$ denotes the cardinality of $\Omega$.  The energy of $v$ in the subspace $S$ is $\|P_S v||_2^2$, where $P_S$ denotes the projection operator onto $S$.  There are two natural estimators of $\|P_S v||_2^2$ based on $v_\Omega$.  The first is simply to form the $n\times 1$ vector $\widetilde v$ with elements $v_i$ if $i \in \Omega$ and zero if $i \not \in \Omega$, for $i=1,\dots,n$.
This `zero-filled' vector yields the simple estimator $\|P_S \widetilde v\|_2^2$.  Filling missing elements with zero is a fairly common, albeit na\"{i}ve, approach to dealing with missing data.  Unfortunately, the estimator $\|P_s \widetilde v\|_2^2$ is fundamentally flawed.  Even if $v \in S$, the zero-filled vector $\widetilde v$ does not necessarily lie in $S$.


A better estimator can be constructed as follows.  Let $U$ be an $n \times r$ matrix whose columns span the $r$-dimensional subspace $S$. Note that for any such $U$, $P_S = U (U^T U)^{-1} U^T$.  With this representation in mind, let $U_\Omega$ denote the $|\Omega| \times r$ matrix, whose rows are the $|\Omega|$ rows of $U$ indexed by the set $\Omega$, arranged in lexigraphic order.  Since we only observe $v$ on the set $\Omega$, another approach to estimating its energy in $S$ is to assess how well $v_\Omega$ can be represented in terms of the rows of $U_\Omega$.  Define the projection operator $P_{S_\Omega} := U_\Omega (U^T_\Omega U_\Omega)^{\dagger}U_\Omega^T$,  where $^\dagger$ denotes the pseudoinverse. It follows immediately that if $v \in S$, then $\|v - P_S v||_2^2 = 0$ and $\|v_\Omega - P_{S_\Omega} v_\Omega\|_2^2 = 0$, whereas $\|\widetilde v - P_S \widetilde v\|_2^2$ can be significantly greater than zero.  This property makes $\| P_{S_\Omega} v_\Omega\|_2^2$ a much better candidate estimator than $\|P_S \widetilde v\|_2^2$.  However,
if $|\Omega| \leq r$, then it
it is possible that 
$\|v_\Omega - P_{S_\Omega} v_\Omega\|_2^2 = 0$,
even if $\|v - P_S v||_2^2 >0$. 
Our main result shows that if $|\Omega|$ is just slightly greater than
$r$, then with high probability $\|v_\Omega - P_{S_\Omega} v_\Omega\|_2^2$
is very close to $\frac{|\Omega|}{n} \|v - P_S v||_2^2$.

\section{Main Theorem}
\label{sec:main}


Let us now focus on our main goal of detecting from a very small number of samples whether there is energy in a vector $v$ outside the $r$-dimensional subspace $S$.  In order to do so, we must first quantify how much information we can expect each sample to provide.  The authors in~\cite{candesrecht} defined the \emph{coherence} of a subspace $S$ to be the quantity
 $$\mu(S) := \frac{n}{r} \max_j \| P_{S} e_j \|_2^2\,.$$  That is, $\mu(S)$ measures the maximum magnitude attainable by projecting a standard basis element onto $S$.   Note that $1 \leq \mu(S) \leq \tfrac{n}{r}$.  The minimum $\mu(S)=1$ can be attained by looking at the span of any $r$ columns of the discrete Fourier transform.  Any subspace that contains a standard basis element will maximize $\mu(S)$.  For a vector $z$, we let $\mu(z)$ denote the coherence of the subspace spanned by $z$.  By plugging in the definition, we have  $$\mu(z) =  \frac{n \|z\|_\infty^2}{\|z\|_2^2}\,.$$
  
To state our main theorem,  write $v=x+y$ where $x \in S$ and $y \in S^{\perp}$. Let the entries of $v$ be sampled uniformly with replacement. Again let $\Omega$ refer to the set of indices for observations of entries in $v$, and denote $|\Omega| = m$.  Given these conventions, we have the following.

\begin{theorem}
Let $\delta > 0$ and $m \geq \frac{8}{3} r\mu(S) \log\left(\frac{2r}{\delta}\right)$. Then with probability at least $1-4\delta$,

\begin{equation*}
 \frac{m(1-\alpha) - r\mu(S)\frac{(1+\beta)^2}{(1-\gamma)}}{n} \|v - P_S v\|_2^2 \leq \|v_\Omega - P_{S_\Omega} v_\Omega \|_2^2 
 \end{equation*} and
 \begin{equation*}
 \|v_\Omega - P_{S_\Omega} v_\Omega \|_2^2 \leq (1+\alpha)\frac{m}{n}\|v - P_S v\|_2^2
\end{equation*} where $\alpha = \sqrt{\frac{2\mu(y)^2}{m}\log\left(\frac{1}{\delta}\right)}$, $\beta = \sqrt{2\mu(y)\log\left(\frac{1}{\delta}\right)}$, and $ \gamma = \sqrt{\frac{8r\mu(S)}{3m} \log\left(\frac{2r}{\delta}\right)}$.
\label{mainthm}
\end{theorem}

\begin{proof}
In order to prove the theorem, we split the quantity of interest into three terms and bound each with high probability. Consider $\|v_\Omega - P_{S_\Omega}v_\Omega\|_2^2 = \|y_\Omega - P_{S_\Omega} y_\Omega \|_2^2$. Let the $r$ columns of $U$ be an orthonormal basis for the subspace $S$. We want to show that 

\begin{equation}
\|y_\Omega - P_{S_\Omega} y_\Omega \|_2^2 = \|y_\Omega\|_2^2 - y_\Omega^T U_\Omega\left(U_\Omega^TU_\Omega\right)^{-1}U_\Omega^Ty_\Omega
\label{term}
\end{equation}
is near $\frac{m}{n} \|y\|_2^2$ with high probability.  To proceed, we need the following three Lemmas whose proofs can be found in the Appendix.


\begin{lemma}
With the same notations as Theorem~\ref{mainthm},
\begin{equation*}
(1-\alpha)\frac{m}{n}\|y\|_2^2 \leq \|y_\Omega\|_2^2 \leq (1+\alpha)\frac{m}{n}\|y\|_2^2
\label{normy}
\end{equation*}
with probability at least $1-2\delta$. \label{normybound}
\end{lemma}

\begin{lemma}
With the same notations as Theorem~\ref{mainthm},
$$ \|U_\Omega^Ty_\Omega\|_2^2 \leq (\beta+1)^2 \frac{m}{n} \frac{r\mu(S)}{n} \|y\|_2^2$$ with probability at least $1-\delta$.
\label{normutybound}
\end{lemma}

\begin{lemma}
With the same notations as Theorem~\ref{mainthm},
$$\|\left(U_\Omega^TU_\Omega\right)^{-1}\|_2 \leq \frac{n}{(1-\gamma)m}$$ 
with probability at least $1-\delta$, provided that $\gamma<1$.
\label{uuinvbound}
\end{lemma}


To apply these three Lemmas, write the second term of Equation (\ref{term}) as
$$ y_\Omega^TU_\Omega\left(U_\Omega^TU_\Omega\right)^{-1}U_\Omega^Ty_\Omega = \|W_\Omega U_\Omega^T y_\Omega\|_2^2$$ where $W_\Omega^TW_\Omega = \left(U_\Omega^TU_\Omega\right)^{-1}$.  By Lemma~\ref{uuinvbound}, $U_\Omega^T U_\Omega$ is invertible under the assumptions of our theorem, and hence $W_\Omega$ is well-defined and has spectral norm bounded by the square root of the inverse of the smallest eigenvalue of $U_\Omega^T U_\Omega$. That is, we have

\begin{eqnarray*}
\|W_\Omega U_\Omega^T y_\Omega\|_2^2 &\leq& \|W_\Omega\|_2^2 \|U_\Omega^T y_\Omega\|_2^2\\
&=& \|W_\Omega^TW_\Omega\|_2  \|U_\Omega^T y_\Omega\|_2^2 \\
&=& \|\left(U_\Omega^TU_\Omega\right)^{-1}\|_2 \|U_\Omega^Ty_\Omega\|_2^2\,.
\end{eqnarray*}
$\|\left(U_\Omega^TU_\Omega\right)^{-1}\|_2$ is bounded by Lemma~\ref{uuinvbound} and $ \|U_\Omega^Ty_\Omega\|_2$ is bounded by Lemma~\ref{normutybound}. Putting these two bounds together with the bounds in Lemma~\ref{normybound} and using the union bound, we have that with probability at least $1-4\delta$
\begin{eqnarray*}
(1+\alpha)^2\frac{m}{n}\|y\|_2^2 \geq \|y_\Omega\|_2^2 - \|\left(U_\Omega^TU_\Omega\right)^{-1}\|_2 \|U_\Omega^Ty_\Omega\|_2^2 \\ 
\geq (1-\alpha)^2 \frac{m}{n}\|y\|_2^2 -  \frac{(\beta+1)^2r\mu(S)}{(1-\gamma)n} \|y\|_2^2 
\end{eqnarray*}
giving us our bound.
\end{proof}

%
%

\section{Discussion and Numerical Experiments}
\label{sec:exp}
In this section we wish to give some intuition for the lower bound in Theorem~\ref{mainthm} and show simulations of the estimate $\|v_\Omega - P_{S_\Omega} v_\Omega\|_2$. If the parameters $\alpha, \beta, \gamma$ are very near $0$, our lower bound is approximately equal to $$\frac{m-r\mu(S)}{n} \|v - P_S v\|_2$$ For an incoherent subspace, the parameter $\mu(S) = 1$. In this case, for $m\leq r$ the bound is $\leq 0$, which is consistent with the fact that $\|v_\Omega - P_{S_\Omega} v_\Omega\|_2 = 0$ always for $m \leq r$. Once $m \geq r+1$, linear algebraic reasoning tells us that $\|v_\Omega - P_{S_\Omega} v_\Omega\|_2$ will be strictly positive with positive probability; Theorem~\ref{mainthm} goes further to say the norm is strictly positive with high probability once $m \sim O(r log r)$. 

%
%

The parameters $\alpha, \beta, \gamma$ all depend on $\sqrt{\log\left(\frac{1}{\delta}\right)}$; these parameters grow as $\delta$ gets very small. Increasing the number of observations $m$ will counteract this behavior for $\alpha$ and $\gamma$, but this does not hold for $\beta$. In fact, even if the vector $y$ is incoherent and $\mu(y)=1$, its minimum value, then $\beta = 2$ for $\delta \approx .135$. To get $\beta$ very near zero, $\delta$ must be \emph{very} near one, but this is not a useful regime. 

We can see, however, that in simulations these large constants are somewhat irrelevant; The large deviations analysis needed for the proof is overly conservative in most cases.

\begin{figure}[!t]
\centerline{\subfloat[Incoherent subspace (random Gaussian basis). $\mu(S)\approx 1.5$, $\mu(y) \approx 13.6$.]{\includegraphics[width=1.5in]{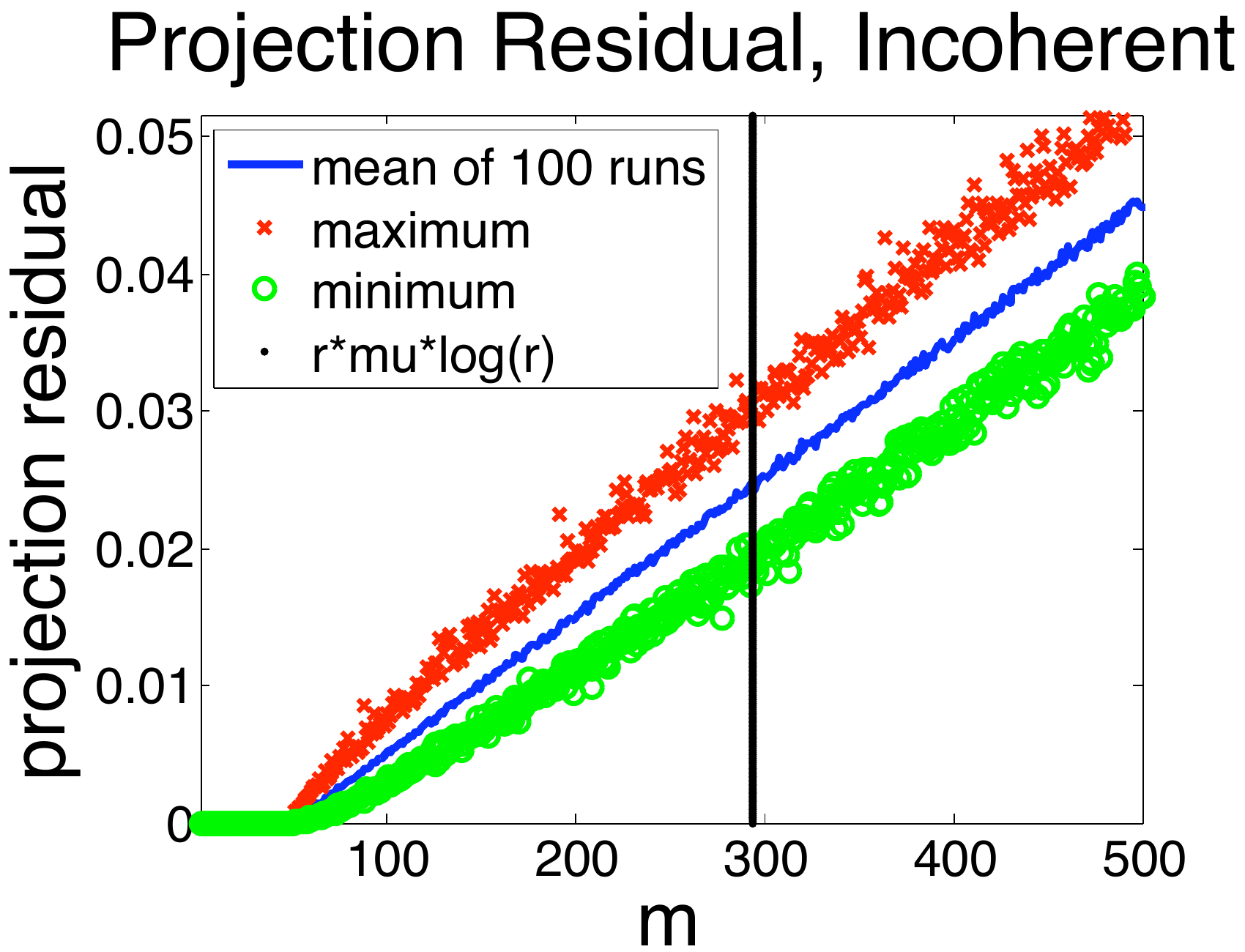}\label{fig_first_case}}
\hfil
\subfloat[Coherent subspace. $\mu(S)\approx4.1$, $\mu(y)\approx47.0$.]{\includegraphics[width=1.5in]{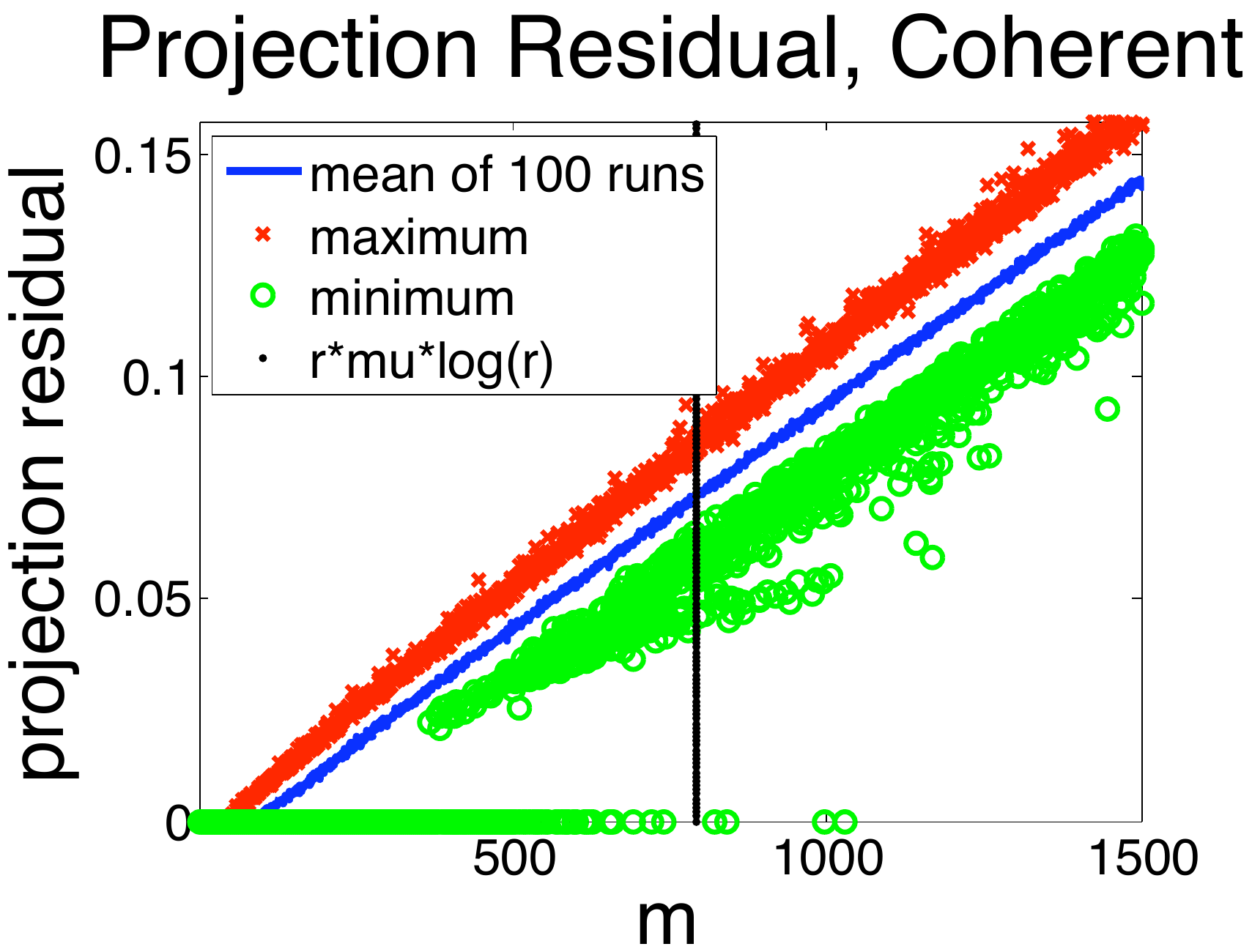}\label{fig_second_case}}}

\caption{These plots show the projection residual $\|v_\Omega - P_{S_\Omega}v_\Omega\|_2^2$  over 100 simulations. Each of the simulations has a fixed subspace, vector $v \in S^\perp$ and sample size $m$, but different sample set $\Omega$ drawn \emph{without} replacement. The problem size is $n=10000$, $r=50$. }
\label{fig_sim}
\end{figure}

%
This plays out in the simulations shown in Figure~\ref{fig_sim}, where we see that for very incoherent subspaces, $ \|v_\Omega - P_{S_\Omega} v_\Omega\|_2 $ is always positive for $m > r \mu(S) \log r$. The plots show the minimum, maximum and mean value of $ \|v_\Omega - P_{S_\Omega} v_\Omega\|_2 $ over 100 simulations, for fixed $S$ and fixed $v$ such that $\|v\|_2^2 = 1$ and $v \in S^\perp$. For each value of the sample size $m$, we sampled 100 different instances of $\Omega$\emph{without} replacement, giving us a realistic idea of how much energy of $v$ is captured by $m$ samples. Our simulations for the Fourier basis and a basis made of orthogonalized Gaussian random vectors always showed the estimate to be positive for  $m > r \mu(S) \log r$, even for the worst-case simulation run. For more coherent subspaces, we often (but not always) see that the norm is positive as long as $m > r\mu(S) \log r$.


\begin{figure}
\centering
\includegraphics[width=1.5in]{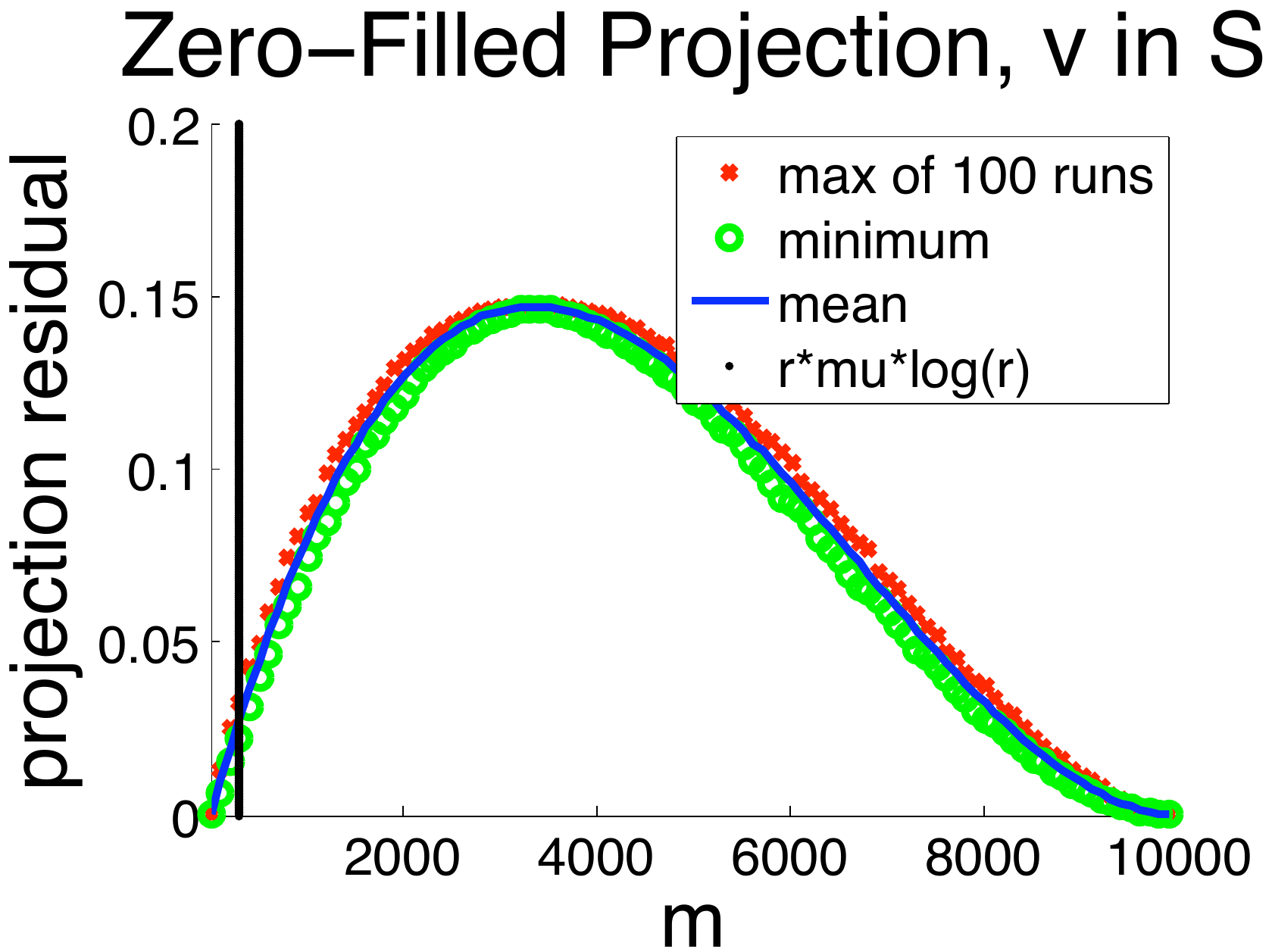}
\caption{Simulation results for the zero-filling approach, $v \in S$, $\|v\|_2^2 = 1$. The basis used is a random Gaussian basis, $r=50$, $n=10000$, $\mu(S)\approx1.5$, $\mu(y)\approx17.9$. 
Note that the zero-filled residuals can be made arbitrarily large by increasing $\|v\|_2^2$.}
\label{fig_zerofill}
\end{figure}

\section{Matched Subspace Detection}
\label{sec:detect}

We have the following detection set up. Our hypotheses are $\mathcal{H}_0 : v \in S$ and
$\mathcal{H}_1 : v \notin S$
%
%
%
and the test statistic we will use is $$t(v_\Omega) = \|v_\Omega - P_{S_\Omega} v_\Omega \|_2^2 \overset{{{\cal H}_1}}{\underset{{{\cal H}_0}}{\gtrless}} \eta$$ 

In the noiseless case, we can let $\eta = 0$; our result in Theorem~\ref{mainthm} shows for $\delta>0$, the probability of detection is $\mathit{P}_D = \mathbb{P}\left[t(v_\Omega) > 0 | \mathcal{H}_1 \right] \geq 1-4\delta$ as long as $m$ is large enough, and we also have that the probability of false alarm is zero, $\mathit{P}_{FA} = \mathbb{P} \left[ t(v_\Omega) > 0 | \mathcal{H}_0 \right] = 0$ since the projection error will be zero when $v \in S$. 

When we introduce noise we have the same hypotheses, but we compute the statistic on $\widetilde{v}_\Omega = v_\Omega + w$ where $w \sim \mathcal{N}(0,1)$ is Gaussian white noise: $$t(\widetilde{v}_\Omega) = \|\widetilde{v}_\Omega - P_{S_\Omega} \widetilde{v}_\Omega \|_2^2 \overset{{{\cal H}_1}}{\underset{{{\cal H}_0}}{\gtrless}} \eta_\lambda$$ We choose  $\eta_\lambda$ to fix the probability of false alarm: $$\mathbb{P} \left[ t(\widetilde{v}_\Omega) > \eta_\lambda | \mathcal{H}_0 \right] \leq \lambda = \mathit{P}_{FA}$$ Then we have from \cite{scharf} that $t(\tilde{v}_\Omega)$ is distributed as a non-central $\chi^2$ with $r$ degrees of freedom and non-centrality parameter $\|v_\Omega - P_{S_\Omega} v_\Omega \|_2^2$, and that $\mathit{P}_D$ is monotonically increasing with the non-centrality parameter. Putting this together with Theorem~\ref{mainthm} we see that as $m$ grows, $\|v_\Omega - P_{S_\Omega} v_\Omega \|_2^2$ grows and thus the probability of detection grows.

We now show why the heuristic approach of zero-filling the incomplete vector $v_\Omega$ does not work.  
As we described in Section~\ref{sec:estimator}, the zero-filling approach is to fill the vector $v$ with zeros and then project onto the full subspace $S$. We denote the zero-filled vector as $v_0$ and then calculate the projection energy only on the observed entries: $$t_0(v_\Omega) = \|v_\Omega - \left( P_S v_0 \right)_\Omega \|_2^2  \overset{{{\cal H}_1}}{\underset{{{\cal H}_0}}{\gtrless}} \eta$$ 
%
%
Simple algebraic consideration reveals that $t_0(v_\Omega)|\mathcal{H}_0$ is positive. In fact, even in the absence of noise, the probability of false alarm can be arbitrarily large as $\|v\|_2^2$ increases.  The value of $t_0(v_\Omega)|\mathcal{H}_0$, based on noiseless observations, is plotted as a function of the number of measurements in Figure~\ref{fig_zerofill}.

We note that for unknown noise power or structured interference, these results can be extended using the GLRT \cite{scharfpap}.

\section{Conclusion}
We have shown that it is possible to detect whether a highly incomplete vector has energy outside a subspace. This is a fundamental result to add to a burgeoning collection of results for incomplete data analysis given a low-rank assumption. Missing data are the norm and not the exception in any massive data collection system, so this result has implications on many other areas of study. 

One of our reviewers shared an insight that the process by which we observe some components and observe erasures in other components can be expressed as a projection operator. It may be possible to extend the results of Theorem~\ref{mainthm} to a wide class of models of random projection operators beyond the class of deletion operators studied here.




\section*{Acknowledgments}
The authors would like to thank the reviewers for their thoughtful comments. This work was supported in part by AFOSR grant FA9550-09-1-0140.

%
%

%
%

\appendix

%
%

\section{Useful Inequalities}

We will need the following two large deviation bounds in the proofs of our Lemmas below.

\begin{theorem}[McDiarmid's Inequality \cite{mcdiarmidcite}] 
Let $X_1, \dots, X_n$ be independent random variables, and assume $f$ is a function for which there exist $t_i$, $i=1, \dots, n$ satisfying $$\sup_{x_1, \dots, x_n, \hat{x_i}} | f(x_1, \dots, x_n) - f(x_1, \dots, \hat{x_i}, \dots, x_n) | \leq t_i$$ where $\hat{x_i}$ indicates replacing the sample value $x_i$ with any other of its possible values. Call $ f(X_1, \dots, X_n) := Y$. Then for any $\epsilon > 0$, 

\begin{equation}
\mathbb{P}\left[ Y \geq \mathbb{E}\left[Y\right] + \epsilon \right] \leq \exp\left(\frac{-2\epsilon^2}{\sum_{i=1}^{n} t_i^2}\right)
\label{mcdiarmideqn1}
\end{equation}

\begin{equation}
\mathbb{P}\left[Y \leq \mathbb{E}\left[Y\right] - \epsilon \right] \leq \exp\left(\frac{-2\epsilon^2}{\sum_{i=1}^{n} t_i^2}\right)
\label{mcdiarmideqn2}
\end{equation}

\label{mcdiarmid}
\end{theorem}

\begin{theorem}[Noncommutative Bernstein Inequality  \cite{Gross09, noncommberncite} ]
Let $X_1, \dots, X_m$ be independent zero-mean square $r \times r$ random matrices. Suppose $\rho_k^2 = max\{\|\mathbb{E}[X_k X_k^T]\|_2, \| \mathbb{E}[X_k^T X_k] \|_2 \}$ and $\|X_k\|_2 \leq M$ almost surely for all $k$. Then for any $\tau > 0$, 

$$\mathbb{P} \left[ \left\| \sum_{k=1}^m X_k \right\|_2 > \tau \right] \leq 2r  \exp \left( \frac{-\tau^2 / 2}{\sum_{k=1}^m \rho_k^2 + M\tau / 3} \right)$$

\label{bernstein}
\end{theorem}

\section{Supporting Lemmas and Proofs}

%
%

We now proceed with the proof of our three central Lemmas.

\begin{proof}[Proof of Lemma~\ref{normybound}]
To prove this we use McDiarmid's inequality from Theorem~\ref{mcdiarmid} for the function $f(X_1, \dots, X_m) = \sum_{i=1}^m X_i$. The resulting inequality is more commonly referred to as  Hoeffding's inequality.

We begin with the first inequality.  Set $X_i = y_{\Omega(i)}^2$. We seek a good value for $t_i$.
Since $y_{\Omega(i)}^2 \leq \|y\|_\infty^2$ for all $i$, we have
\begin{eqnarray*}
\left| \sum_{i=1}^m X_i - \sum_{i\neq k} X_i - \hat{X_k}\right| &=& \left| X_k - \hat{X_k} \right| \leq 2\|y\|_\infty^2
\end{eqnarray*}
We calculate $\mathbb{E}\left[\sum_{i=1}^m X_i \right]$ as follows. Define $\mathbb{I}_{\{\}}$ to be the indicator function, and assume that the samples are taken uniformly with replacement.

\begin{eqnarray*}
\mathbb{E}\left[\sum_{i=1}^m X_i \right] & =&  \mathbb{E}\left[\sum_{i=1}^m y_{\Omega(i)}^2 \right]\\
& = & \sum_{i=1}^m \mathbb{E}\left[ \sum_{j=1}^n y_j^2 \mathbb{I}_{\{\Omega(i)=j\}} \right]
 =  \frac{m}{n} \|y\|_2^2
 \end{eqnarray*}
Plugging into Equation (\ref{mcdiarmideqn2}), the left hand side is

$$ \mathbb{P}\left[ \sum_{i=1}^m X_i \leq \mathbb{E}\left[\sum_{i=1}^m X_i\right] - \epsilon \right] = \mathbb{P}\left[\sum_{i=1}^m X_i \leq  \frac{m}{n} \|y\|_2^2 - \epsilon \right]$$
and letting $\epsilon = \alpha \frac{m}{n} \|y\|_2^2$, we then have that this probability is bounded by $$\exp \left( \frac{-2\alpha^2\left(\frac{m}{n}\right)^2  \|y\|_2^4}{4 m \|y\|_{\infty}^4} \right)$$ Thus, the resulting probability bound is

\begin{equation}
\mathbb{P}\left[\|y_\Omega\|_2^2 \geq (1-\alpha)\frac{m}{n} \|y\|_2^2 \right] \geq 1-\exp\left(\frac{-\alpha^2 m \|y\|_2^4}{2 n^2 \|y\|_{\infty}^4}\right)
\end{equation}
Substituting our definitions of $\mu(y)$ and $\alpha$ shows that the lower bound holds with probability at least $1-\delta$. The argument for the upper bound is identical after replacing Equation (\ref{mcdiarmideqn1}) instead of (\ref{mcdiarmideqn2}).  The Lemma now follows by applying the union bound. \end{proof}

%
%

\begin{proof}[Proof of Lemma~\ref{normutybound}]
We use McDiarmid's inequality in a very similar fashion to the proof of Lemma~\ref{normybound}. Let $X_i = y_{\Omega(i)}U_{\Omega(i)}$, where $\Omega(i)$ refers to the $i^{th}$ sample index. Thus $y_{\Omega(i)}$ is a scalar, and the notation $U_{\Omega(i)}$ refers to an $r \times 1$ vector representing the transpose of the $\Omega(i)^{th}$ row of $U$. 

Let our function $f(X_1, \dots, X_m) = \|\sum_{i=1}^{m}X_i\|_2 = \|U_\Omega^T y_\Omega\|_2$. To find the $t_i$ of the theorem we first need to bound $\|X_i\|$ for all $i$. Observe that $\|U_{\Omega(i)}\|_2 = \|U^T e_i\|_2 = \|P_S e_i\|_2 \leq \sqrt{r\mu(S)/n}$ by assumption. Thus,

$$\|X_i\|_2 \leq |y_{\Omega(i)}| \|U_{\Omega(i)}\|_2 \leq \|y\|_{\infty} \sqrt{r\mu(S)/n}$$
Then observe $\left| f(X_1, \dots, X_m) - f(X_1, \dots, \hat{X_k}, \dots, X_m)\right|$ is

\begin{eqnarray*}
\left| \left\| \sum_{i=1}^{m} X_i \right\|_2 - \left\|\sum_{i\neq k} X_i + \hat{X_k}\right\|_2\right|
& \leq & \left\| X_k - \hat{X_k} \right\|_2 \\
& \leq & \left\| X_k \right\|_2 + \| \hat{X_k} \|_2 \\&\leq& 2\|y\|_{\infty} \sqrt{\frac{r\mu(S)}{n}}\,.
\end{eqnarray*}
Here, the first two inequalities follow from the triangle inequality.
Next we calculate a bound for $\mathbb{E}\left[f(X_1,\dots,X_m)\right] = \mathbb{E}\left[ \left\| \sum_{i=1}^{m} X_i \right\| \right]$. Assume again that the samples are taken uniformly with replacement. We have $$\sum_{k=1}^r U_{jk}^2 = \|P_S e_j \|^2 \leq \frac{r}{n}\mu(S)\,,$$ from which we can see that

%
\begin{eqnarray}
\mathbb{E}\left[ \left\| \sum_{i=1}^{m} X_i \right\|_2^2 \right] & = & \mathbb{E}\left[ \left\| U_\Omega^T y_\Omega \right\|_2^2 \right] \nonumber \\
& = & \sum_{k=1}^r \mathbb{E}\left[  \sum_{i=1}^m \sum_{j=1}^n U_{jk}^2 y_j^2 \mathbb{I}_{\{\Omega(i)=j\}}\right] \label{crosstermscancel}\\
& = & \sum_{k=1}^r m \left(\sum_{j=1}^n U_{jk}^2 y_j^2\right) \frac{1}{n} \label{sampunif}\\
&\leq&\frac{m}{n} \frac{r\mu(S)}{n}  \|y\|_2^2 \nonumber
\end{eqnarray}
The step (\ref{crosstermscancel}) follows because the cross terms cancel by orthogonality. The step (\ref{sampunif}) is because of our assumption that sampling is uniform with replacement. 

Since  $\mathbb{E}\left[\|X\|_2\right] \leq \mathbb{E}\left[\|X\|_2^2\right]^{1/2}$ by Jensen's inequality, we have that $\mathbb{E}\left[ \left\| \sum_{i=1}^{m} X_i \right\|_2 \right] \leq \sqrt{\frac{m}{n}} \sqrt{\frac{r\mu(S)}{n}} \|y\|_2$. Letting $\epsilon = \beta \sqrt{\frac{m}{n}} \sqrt{\frac{r\mu(S)}{n}}  \|y\|_2$ and plugging into Equation (\ref{mcdiarmideqn1}), we then have that the probability is bounded by $$\exp\left(\frac{-2\beta^2\frac{m}{n} \frac{r\mu(S)}{n}  \|y\|_2^2}{4 m \|y\|_{\infty}^2 \frac{r\mu(S)}{n}}\right)$$ Thus, the resulting probability bound is

\begin{equation*}
\mathbb{P}\left[  \|U_\Omega y_\Omega\|_2^2 \geq (1+\beta)^2\frac{mr\mu(S)}{n^2}  \|y\|_2^2 \right] \leq \exp\left(\frac{-\beta^2 \|y\|_2^2}{2n \|y\|_{\infty}^2} \right)
\end{equation*}
%
Substituting our definitions of $\mu(y)$ and $\beta$ shows that the lower bound holds with probability at least $1-\delta$,
completing the proof.\end{proof}

%
%

\begin{proof}[Proof of Lemma~\ref{uuinvbound}]
We use the Noncommutative Bernstein Inequality as follows. Let $X_k = U_{\Omega(k)}U_{\Omega(k)}^T - \frac{1}{n} I_r$, where the notation $U_{\Omega(k)}$ is as before, i.e. is the transpose of the $\Omega(k)^{th}$ row of $U$, and $I_r$ is the $r \times r$ identity matrix. Note that this random variable is zero mean.

We must compute $\rho_k^2$ and $M$. Since $\Omega(k)$ is chosen uniformly with replacement, the $X_k$ are identically distributed, and $\rho$ does not depend on $k$. For ease of notation we will denote $U_{\Omega(k)}$ as $U_k$.

Using the fact that for positive semi-definite matrices, ${\|A-B\|_2 \leq \max\{\|A\|_2,\|B\|_2\}}$, and recalling again that $\|U_{k}\|^2_2 = \|U^T e_k\|^2_2 = \|P_S e_k\|^2_2 \leq r\mu(S)/n$, we have 
$$\left\| U_k U_k^T - \frac{1}{n} I_r \right\|_2 \leq \max\left\{ \frac{r\mu(S)}{n}, \frac{1}{n} \right\}$$
and we let $M := r\mu(S) / n$.

For $\rho$, we note
\begin{eqnarray*}
\left\| \mathbb{E} \left[ X_kX_k^T \right] \right\|_2
& = & \left\| \mathbb{E} \left[ X_k^T X_k \right] \right\|_2 \\
& = & \left\| \mathbb{E} \left[ \left(U_k U_k^T - \frac{1}{n} I_r \right)^2  \right] \right\|_2 \\
& = & \left\| \mathbb{E} \left[ U_k U_k^T U_k U_k^T - \frac{2}{n} U_k U_k^T + \frac{1}{n^2} I_r \right] \right\|_2 \\
& = & \left\| \mathbb{E}\left[U_k U_k^T U_k U_k^T\right] -  \frac{1}{n^2} I_r  \right\|_2 \\
& \leq & \max\left\{\left\| \mathbb{E}\left[U_k U_k^T U_k U_k^T\right]\right\| ,  \frac{1}{n^2}\right\} \\
 &\leq& \max\left\{\frac{ r \mu(S)}{n} \|E[U_k U_k^T]\|_2, \frac{1}{n^2}\right\}\\
& =&\max \left\{ \frac{r \mu(S)}{n^2} \|I_r\|_2,  \frac{1}{n^2}\right\}\\
& =& \frac{r \mu(S) }{n^2}\,.
\end{eqnarray*} 
Thus we let $\rho^2 := r\mu(S) / n^2$.

Now we can apply the Noncommutative Bernstein Inequality, Theorem~\ref{bernstein}. First we restrict $\tau$ to be such that $M\tau \leq m \rho^2$ to simplify the denominator of the exponent. Then we get that

$$2r \exp \left( \frac{-\tau^2 / 2}{m \rho^2 + M\tau/3}\right) \leq 2r \exp \left( \frac{-\tau^2 / 2}{\frac{4}{3} m \frac{r\mu(S)}{ n^2}}\right)$$ and thus

$$\mathbb{P} \left[ \left\| \sum_{k \in \Omega} \left(U_k U_k^T - \frac{1}{n} I_r \right) \right\| > \tau \right] \leq 2r \exp \left( \frac{-3 n^2 \tau^2}{8 mr\mu(S)}\right)$$
Now take $\tau = \gamma m/n$ with $\gamma$ defined in the statement of Theorem~\ref{mainthm}. Since $\gamma<1$ by assumption, $M\tau \leq m \rho^2$ holds and we have

\begin{equation*}
\mathbb{P} \left[ \left\| \sum_{k \in \Omega} \left(U_k U_k^T - \frac{1}{n} I_r \right) \right\|_2 \leq \frac{m}{n} \gamma \right]  \geq  1 - \delta
\end{equation*}
We note that $\left\| \sum_{k \in \Omega} U_k U_k^T - \frac{m}{n} I_r \right\|_2 \leq  \frac{m}{n}  \gamma$ implies that the minimum singular value of $\sum_{k \in \Omega} U_k U_k^T$ is at least $(1-\gamma)\frac{m}{n}$. This in turn implies that

$$\left\|\left(\sum_{k \in \Omega} U_k U_k^T\right)^{-1}\right\|_2 \leq \frac{n}{(1-\gamma)m}$$
which completes the proof. \end{proof}



%
\bibliographystyle{abbrv}
\bibliography{isit2010}



\end{document}